\documentclass{article}

\usepackage{fullpage}
\usepackage[english]{babel}

\usepackage{amsmath}
\usepackage{authblk}
\usepackage{graphicx}
\usepackage{hyperref}
\hypersetup{
     colorlinks   = true,
     allcolors    = blue,
}
\usepackage{pdfcomment}
\usepackage{algorithm, algpseudocode}
\usepackage{mathtools}

\newcommand{\eps}{\varepsilon}
\newcommand{\del}{\delta}
\newcommand{\epsdel}{\left(\eps,\del\right)}
\newcommand{\pair}[1]{\left<#1\right>}

\title{Publishing Wikipedia usage data with strong privacy guarantees}
\date{}

\author[1]{Temilola Adeleye}
\author[2]{Skye Berghel}
\author[2]{Damien Desfontaines}
\author[2]{Michael Hay}
\author[1]{\\Isaac Johnson}
\author[1]{Cléo Lemoisson}
\author[2]{Ashwin Machanavajjhala}
\author[2]{Tom Magerlein}
\author[1]{\\Gabriele Modena}
\author[2]{David Pujol}
\author[2]{Daniel Simmons-Marengo}
\author[1]{Hal Triedman}
\affil[1]{Wikimedia Foundation -- \href{mailto:htriedman@wikimedia.org}{htriedman@wikimedia.org}}
\affil[2]{Tumult Labs -- \href{mailto:science@tmlt.io}{science@tmlt.io}}

\begin{document}

\maketitle

\begin{abstract}
For almost 20 years, the Wikimedia Foundation has been publishing statistics about how many people visited each Wikipedia page on each day.
This data helps Wikipedia editors determine where to focus their efforts to improve the online encyclopedia, and enables academic research.
In June 2023, the Wikimedia Foundation, helped by Tumult Labs, addressed a long-standing request from Wikipedia editors and academic researchers: it started publishing these statistics with finer granularity, including the country of origin in the daily counts of page views.
This new data publication uses differential privacy to provide robust guarantees to people browsing or editing Wikipedia.
This paper describes this data publication: its goals, the process followed from its inception to its deployment, the algorithms used to produce the data, and the outcomes of the data release.
\end{abstract}

\section{Introduction}\label{sec:intro}

Wikipedia and other projects supported by the Wikimedia Foundation are among the most used online resources in the world, garnering hundreds of billions of visits each year from around the world.
As such, the Foundation has access to terabytes of data about visits to a page on a Wikimedia project.
This is called \emph{pageview} data in this document.

The Foundation has been publishing statistics about this data for almost 20 years, through the \emph{Pageview API}~\cite{wmfdownloads}.
This data helps Wikipedia editors measure the impact of their work, and focus their efforts where they are most needed.
Pageview data is also a rich resource for academic research: it has been used to better understand many topics, ranging from user behavior~\cite{wmfresearch} and browsing patterns~\cite{wmfclickstream} to information dissemination~\cite{wmfstudies}, epidemiology~\cite{bragazzi2017global}, online harassment~\cite{wulczyn2017ex}, and others. 
Over time, the Wikimedia Foundation received a number of requests to make these statistics more granular, and publish pageview counts \emph{by country}, to make it even more useful to Wikipedia editors, and enable further academic research.

Addressing such requests for more granular data is aligned with the Foundation's open access policy~\cite{wmfopenaccess}, which seeks to provide as much transparency as possible about how Wikimedia projects operate.
However, the Foundation also considers privacy to be a key component of the free knowledge movement: there cannot be creation or consumption of free knowledge without a strong guarantee of privacy.
These guarantees are expressed by the Foundation’s strict privacy policy~\cite{wmfprivacy} and data retention guidelines~\cite{wmfretention}, which govern how the infrastructure underlying Wikipedia works.
Concretely, people browsing Wikipedia may expect their behavior on the website to stay private: is is crucial to prevent motivated actors to combine this data with outside other data sources in order to spy on or persecute Wikipedia users for their view history, edit history, or other behavior.
It is well-known that simply aggregating data is not, on its own, enough to prevent re-identification risk~\cite{xu2017trajectory,desfontainesblog20210526,cohen2022attacks,dick2023confidence}, so publishing data with a finer geographic granularity warrants an approach with rock-solid privacy guarantees for Wikipedia users and editors.

Differential privacy~\cite{dwork2006calibrating} (DP) provides a way of easing this tension: it allows organizations to both lower and more fully understand the risks of releasing data.
Therefore, the Wikimedia Foundation decided to investigate the use of differential privacy to release daily pageview data, sliced by country.
After an in-depth comparison of available open-source tools~\cite{wmfdpcomparison}, the Wikimedia Foundation decided to use Tumult Analytics~\cite{tumultanalyticssoftware,tumultanalyticswhitepaper} and started a collaboration with Tumult Labs to design and deploy a DP pipeline for this data release.
The pipeline is now deployed, and the published data provides useful insights to anyone interested in better understanding Wikipedia usage.

This document describes this data release in more detail.
\begin{itemize}
    \item In Section~\ref{sec:workflow}, we present the high-level workflow that we followed towards the deployment of a differentially private data release.
    \item In Section~\ref{sec:problem}, we outline the problem statement and the success metrics for this data release.
    \item In Section~\ref{sec:algorithms}, we describe the technical algorithms used for this data release.
    \item In Section~\ref{sec:outcomes}, we summarize the results of this deployment.
\end{itemize}

\section{High-level workflow for differential privacy deployments}\label{sec:workflow}

\begin{figure}[ht!]
\centering
\pdftooltip{\includegraphics[width=0.8\linewidth]{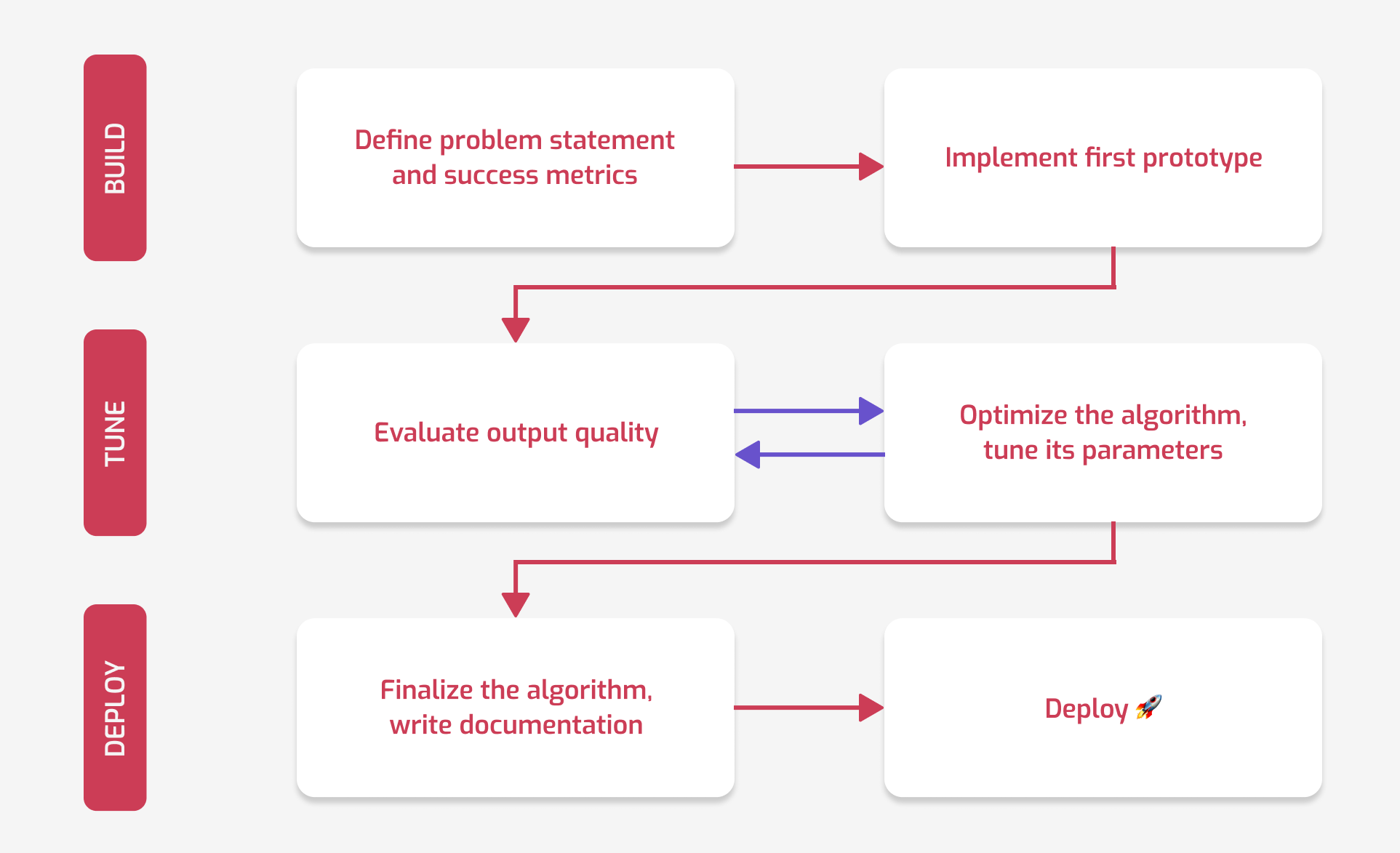}}{
A diagram with three rows of two boxes with different colors.
The top row, labeled “build”, starts from “Define problem statement and success metrics”, and an arrow points to “Implement first prototype”.
An arrow from there goes to the second row, labeled “Tune”, starting with “Evaluate output quality”.
Arrows in both directions go to the second box of the second row, “Optimize the algorithm, tune its parameters”.
An arrow goes to this box to the third row, labeled “Deploy”, to a box “Finalize the algorithm, write documentation”, and a last arrow goes to “Deploy”.}
\caption{\label{fig:workflow}A standardized workflow for differentially private data releases.}
\end{figure}

The process to launch a DP data product follows a standard workflow, with three main stages: \emph{Build}, \emph{Tune}, and \emph{Deploy}. 
The entire process is outlined in Figure~\ref{fig:workflow}; its three main stages are as follows.
\begin{enumerate}
\item In the initial Build stage, the goal is to gain a good understanding of the problem and its requirements, and implement a first-cut algorithm. There are two steps in this initial stage.
First, we properly define the problem, and determine what success looks like for this project. This involves talking to stakeholders to understand what the data will be used for, and what accuracy metrics capture downstream use cases well.
Second, we build a prototype mechanism. This is a first rough attempt at solving the data release problem, and it exposes the ``levers'' inherent to the project. Which choices did we have to make while building the prototype? Which of these choices can later be modified to pick different trade-offs between utility or privacy metrics?
\item Then, in the Tune step, we use these levers to experiment with different settings and optimize the algorithm. Using the success metrics defined in the previous step, we iteratively evaluate and adjust the algorithm, making changes until it produces data that is fit for use and satisfies the privacy requirements.
\item Finally, in the Deploy stage, we finalize the algorithm, obtain the necessary approvals to publish the data, write documentation about the data publication mechanism for future data users and pipeline maintainers, and deploy it in production.
\end{enumerate}

In Section~\ref{sec:problem}, we outline the output of the very first step: the definition of problem statement and its success metrics. 
Then, in Section~\ref{sec:algorithms}, we will describe the output of the Tune stage: what does the final algorithm look like, after the multiple rounds of iteration on the initial prototype.

\section{Problem statement and success metrics}\label{sec:problem}

In this Section, we describe the desired output data (Section~\ref{sec:desired-output}), the schema and characteristics of the input data (Section~\ref{sec:input}), the privacy goals of this data release (Section~\ref{sec:privacy-goal}), and the accuracy metrics used to quantify success (Section~\ref{sec:accuracy}).

\subsection{Desired output data}\label{sec:desired-output}

The pre-existing Pageview API publishes data about the number of times each Wikimedia page was visited during a given day. Each page is identified by two fields:
\begin{itemize}
    \item its \emph{project}, e.g. \texttt{fr.wikipedia} (the French-language version of Wikipedia), \texttt{zh.wikibooks} (the Chinese version of Wikibooks, an open-content textbook collection), \texttt{wikidata} (a central storage for structured data), etc.;
    \item its \emph{page ID}, a numeric identifier uniquely identifying each page within a project.
\end{itemize}
Table~\ref{tab:pageview-api} is a fictitious sample of the kind of data available via the Pageview API. For example, the first line indicates that there were 4217 visits to the page with ID 23110294 on the English version of Wikipedia on April 2nd, 2023.

\begin{table}[h]
    \centering
    \begin{tabular}{l|l|l|r}
        Project & Page ID & Date & Count \\\hline
        en.wikipedia & 23110294 & 2023-04-02 & 4217\\
        fr.wikipedia & 28278 & 2023-04-02 & 710 \\
        \dots & \dots & \dots & \dots
    \end{tabular}
    \caption{\label{tab:pageview-api}A fictitious sample from the data made publicly available via the Pageview API.}
\end{table}

The goal of this project is to publish more granular data, and also release daily pageview counts \emph{per country}.
A fictitious sample of the desired output data appears in Table~\ref{tab:desired-output}.
For example, the first line indicates that 92 of the previously-mentioned visits originated from Switzerland.

\begin{table}[h]
    \centering
    \begin{tabular}{l|l|l|l|r}
        Project & Page ID & Date & Country & Count \\\hline
        en.wikipedia & 23110294 & 2023-04-02 & CH & 92\\
        fr.wikipedia & 28278 & 2023-04-02 & FR & 101 \\
        \dots & \dots & \dots & \dots
    \end{tabular}
    \caption{\label{tab:desired-output}A fictitious sample from the data that we would like to publish as part of this project.}
\end{table}

\subsection{Input data}\label{sec:input}

This project uses two input datasets: the \emph{current pageviews dataset}, and the \emph{historical pageviews dataset}.

\paragraph{Current pageviews dataset}
As users visit the site, their individual pageviews are recorded and stored in the current pageviews dataset.
This dataset contains all pageviews across all Wikimedia projects for the last 90 days.
Because of the Wikimedia Foundation's commitment to minimal data retention, this data is only kept in this form for 90 days.
Table~\ref{tab:current-data} is a fictitious sample of the current pageviews dataset, showing only the columns of interest for this project: project, page ID, date and time, and country.

\begin{table}[h]
    \centering
    \begin{tabular}{c|c|c|c}
        Project & Page ID & Date and Time & Country \\\hline
        en.wikipedia & 23110294 & 2023-04-02 10:32:45 & CH \\
        fr.wikipedia & 28278 & 2023-04-02 18:53:11 & FR \\
        \dots & \dots & \dots & \dots
    \end{tabular}
    \caption{\label{tab:current-data}A fictitious sample of the columns of interest from the current pageviews dataset.}
\end{table}

Note that in contrast to similar logging infrastructure for most websites, this data does not contain a persistent user identifier. 
Most visits to Wikimedia projects come from logged-out users, and the Wikimedia Foundation intentionally did not implement a user tracking mechanism which would provide a cookie ID and allow the Foundation's systems to recognize whether two records came from the same user.
This practice is good for data minimization, but it makes it more difficult to obtain user-level differential privacy guarantees, which requires bounding the number of contributions coming from the same user.
We come back to this challenge in Section~\ref{sec:client-side}.

\paragraph{Historical pageviews dataset}
Past the initial 90-day retention period, pageviews are aggregated as hourly totals, broken down by project, page id, country, and a number of user characteristics.
These aggregates are then stored in the historical pageviews dataset.
Table~\ref{tab:historical-data} is a fictitious sample of the historical pageviews dataset, again showing only the columns of interest.


\begin{table}[h]
    \centering
    \begin{tabular}{c|c|c|c|c}
         Project & Page ID & Date and Time & Country & Count \\\hline
        en.wikipedia & 23110294 & 2023-04-02 10:00 & CH & 11 \\
        fr.wikipedia & 28278 & 2023-04-02 18:00 & FR & 15 \\
        \dots & \dots & \dots & \dots & \dots
    \end{tabular}
    \caption{\label{tab:historical-data}A fictitious sample of the columns of interest from the pre-aggregated historical pageviews dataset.}
\end{table}

This pre-aggregated data also poses a challenge for performing DP calculations: it is not possible to determine which contributions came from which users, and therefore to bound the number contributions coming from each user.

\subsection{Privacy goal}\label{sec:privacy-goal}

When using differential privacy, one has to decide what to protect in the data; or, equivalently, what the definition of the neighboring databases should be.
For long-running pipelines that publish data regularly over an unbounded time period, there are two aspects to this choice: what are the intervals of time considered as part of the unit of privacy, and what are we protecting in each of these intervals.
Then, a follow-up question is the choice of privacy parameters: the numeric value of $\eps$ and $\del$.

Our goal is to publish data daily: it is natural to use a daily time period in the unit of privacy.
This interval is consistent with almost all other long-running DP deployments, like Apple's telemetry collection, or Google's and Meta's data releases related to the COVID-19 crisis.
Other releases use a shorter period, like Microsoft's telemetry in Windows.
There is no overlap between days: the privacy parameters for each user-day are fixed and do not increase over time.

This choice of unit of privacy means that if a user were to regularly visit the same page from the same country across multiple devices (or clearing their cookies between each page visit) over a long period of time, this behavior could potentially be observed in the output data.
Another caveat is that this data release surfaces group-level trends, like minority language group activity on Wikimedia projects within a country.
These insights can be helpful (e.g. allow for dedicated support to that minority language group) but could also carry risks (e.g. by causing government persecution of this minority group).
We mitigate these risks by choosing conservative privacy parameters, which translate to a reasonable level of protection over longer time periods, by holding off on releasing data for certain countries, and by only releasing aggregates that are above a certain threshold.

Protecting each individual Wikipedia user during each day is impossible to achieve entirely without a way to link people's identities across records and devices.
Because the Wikimedia Foundation does not have nor want the capability to link records in such a way, we instead attempt to protect the contribution of each \emph{device} during each day.
For the data based on the current pageviews dataset, we achieve this goal using \emph{client-side contribution bounding}, as described in Section~\ref{sec:client-side}.
For the data based on the historical pageviews dataset, we cannot bound user contributions.
Instead, we choose to protect a fixed number of daily pageviews, denoted by $m$.
This provides an equivalent level of protection to users who contribute fewer than $m$ pageviews per day.
Users who contribute more than $m$ pageviews will incur a larger privacy loss, proportional to the amount by which their contributions exceed $m$.
This number is set to 300 for data prior to February 8th, 2017, and to 30 for data between February 9th, 2017 to February 5th, 2023.
Table~\ref{tab:privacy-units} summarizes the privacy units chosen for this project.

\begin{table}[htpb]
    \centering
    \begin{tabular}{c|c}
         Time period of the input data & Unit of privacy \\\hline
         July 1st, 2015 -- February 8th, 2017 & 300 daily pageviews \\
         February 9th, 2017 -- February 5th, 2023 & 30 daily pageviews \\
         February 6th, 2023 -- present & one user-day \\
    \end{tabular}
    \caption{\label{tab:privacy-units}A summary of the privacy units used in this project.}
\end{table}

This difference in how many contributions we protect is due the fact that in February 2017, a change occurred to the way the input data was generated.
Prior to February 8th, 2017, users who were editing a Wikimedia page and used the Web UI to preview their changes were recorded as one pageview each time the preview refreshed.
This meant that during a lengthy editing session, an editor could plausibly rack up many pageviews on the same page.
When combined with our inability to limit user contributions, this created a markedly different risk level before/after this date, that our historical pageviews algorithm had to address.
Starting on February 9th, 2017, previews were no longer recorded as pageviews. 

For privacy parameters, we use zero-concentrated DP~\cite{bun2016concentrated} with $\rho=0.015$\footnote{Which is a strictly stronger guarantee than $\epsdel$-DP~\cite{dwork2006our} with $\eps=1$ and $\del=10^{-7}$.} for the more recent data, and pure DP with $\eps=1$ for the historical data.
These values are generally considered to be conservative among differential privacy researchers and practitioners~\cite{near2022differential}, and are lower than most practical DP deployments~\cite{desfontainesblog20211001}.

\subsection{Accuracy metrics}\label{sec:accuracy}

We measure utility along three dimensions: the \emph{relative error distribution}, the \emph{drop rate}, and the \emph{spurious rate}.
Each of these metrics is computed using the \emph{true data} as a baseline: the data that corresponds to simply running a group-by query (either counting the number of rows, for the current pageviews dataset, or summing the counts, for the historical pageviews dataset), without any contribution bounding, noise addition, nor suppression.

\paragraph{Relative error distribution}

We are releasing pageview counts, and the DP process will inject statistical noise into these counts.
Thus, it is natural to want to measure how much noise is added to these counts.
We measure accuracy according to \emph{relative error}: the relative error of each noisy count $\hat{c}$ is $\left|\hat{c}/c\right|$, where $c$ is the true count.
Of course, we are releasing many counts, so we need to look at the \emph{distribution} of relative error.
More specifically, we look at the percentage of released counts having a relative error smaller than $10\%$, $25\%$, and $50\%$.

\paragraph{Drop rate}

The DP algorithm uses \emph{suppression}: if a noisy count is lower than a given threshold, we remove it from the output data.
To quantify the data loss due to this suppression step, it is natural to compute the \emph{drop rate}: the percentage of counts that do not appear in the output, even though they were non-zero in the true data.
In the true data, however, many of the counts are very low; suppressing such counts is not as bad as suppressing a popular page.
Therefore, we compute the percentage of pages that were suppressed among pages whose true counts is larger than a fixed threshold $t$ (the \emph{drop rate above $t$}), as well as the percentage of pages that were suppressed among the top 1000 rows in the true data (the \emph{top-1000 drop rate}).

\paragraph{Spurious rate}

Many page, project, and country combinations receive zero pageviews on any particular day.
When noise is added to these zero counts, it is likely that they will end up with positive (though comparatively small) counts. 
We refer to these as \emph{spurious} counts. 
Spurious counts can mislead data users by wrongly indicating that some combinations had activity.
They also increase the size of the output dataset, which can pose a usability challenge.
Therefore, we compute an additional metric: the \emph{spurious rate}, which captures the ratio of spurious counts among all counts that appear in the output.
\section{Technical description of the algorithms}\label{sec:algorithms}

In this Section, we describe the algorithms used to generate the differentially private data.
For simplicity, we refer to a $<$page ID, project$>$ pair as a \emph{page}.

\subsection{Current pageviews}\label{sec:current}

For the data using the current pageviews dataset, we want to provide privacy guarantees that protect each user during each day.
This requires bounding the maximum number of pageviews that each user can contribute during a single day.
The typical way to perform such contribution bounding is to use a user identifier to sub-sample the number of contributions from each user, taking the first $k$ records~\cite{korolova2009releasing}, or using reservoir sampling~\cite{wilson2020differentially}.
However, without a user identifier, we had to use a novel and alternative approach to this problem: \emph{client-side filtering}. 

\subsubsection{Client-side filtering}\label{sec:client-side}

Without user IDs, the server cannot know whether multiple contributions come from the same user, and perform contribution bounding to get user-level privacy guarantees.
Instead, we add some logic to the client side.
Each end-user device counts their number of contributions logged in each day, and sends each contribution along with a boolean flag, indicating whether this contribution should be used in the server-side DP computation. 
The criteria used for inclusion in the input to the DP algorithm is as follows: each day, we include the first 10 \emph{unique} pageviews.
This means that if a user visits the same page multiple times in a day, only the first visit will be counted.
This also means that if a user visits more than 10 distinct pages in a day, all pageviews after the 10th visits will not be included.

Pseudocode for this client-side filtering step can be found in Algorithm~\ref{algo:client-side}.
Note that this algorithm does not keep track of the raw page IDs in the client-side cookie.
Instead, it uses a salted hash function~\cite{wmfsalt} to remember which page IDs were already visited.
This provides an additional level of protection against an attacker that would obtain access to this cookie.

\begin{algorithm}[h]
\caption{Client-side filtering algorithm}
\begin{algorithmic}[1]
\Require $P = p_1, p_2, \dots$: a stream of pageviews.
\Require $H$: a salted hash function
\Require $k$: the number of unique pageviews to include.
\Ensure The output is a stream of the same pageviews, each one annotated with a boolean indicating whether it should be included in the DP computation. This boolean is $\texttt{true}$ iff the pageview comes from a page not output before, and 
\State $S \leftarrow \{\}$
\For{$p$ in $P$} 
    \If{$\left|S\right|\ge k$ or $H(p) \in S$}
        \State Output $\left<p, \texttt{false}\right>$
    \Else
        \State $S \leftarrow S \cup {H(p)}$
        \State Output $\left<p, \texttt{true}\right>$
    \EndIf
\EndFor
\end{algorithmic}
\label{algo:client-side}
\end{algorithm}

Client-side filtering upholds the Wikimedia Foundation's data minimization principle: only the absolute minimal information needed to perform the contribution bounding --- a boolean value associated with each pageview to indicate whether it should be counted --- is added to the logging infrastructure.
Alternatives such as using identifiers or a counter that increments for each contribution would have required sending more data to the server, and increase fingerprinting risk.

\subsubsection{Server-side algorithm}\label{sec:server-side}

Once each pageview was annotated by the client-side filtering algorithm, it is used as input in a server-side differentially private algorithm.
This algorithm, run daily on the data from the previous day, has three stages.
\begin{enumerate}
    \item First, we collect the list of $<$page, country$>$ tuples to aggregate over.
    \item Second, we count the number of pageviews in each group, and we add noise to each count.
    \item Finally, we suppress low counts, and publish the data.
\end{enumerate}

The list of all possible tuples is, in theory, known in advance: the list of Wikimedia pages and countries are both public information.
However, the majority of $<$page, country$>$ combinations do not appear in the input data: including all of them would be inefficient and lead to increased spurious data.
Instead, we use existing public data to only include a small fraction of these possible counts.
On each day, we list all Wikimedia pages with more than $t$ global pageviews, according to the existing Pageview API, where $t$ is an arbitrary ingestion threshold.
Then, we take the cross-product between these pages and the list of countries\footnote{This list is based on~\cite{wmfcountries}; excluding countries identified by the Wikimedia Foundation as potentially dangerous for journalists or internet freedom~\cite{wmfcountryprotectionlist}.} to create the groups.

The second step uses the Gaussian mechanism~\cite{dwork2014algorithmic} to add noise to counts.
This provides two advantages.
First, because each user can contribute to at most 10 \emph{different} $<$page, country$>$ tuples, but only once to each, we get a tighter $L_2$ sensitivity bound ($\sqrt{k}$) than if we had used $L_1$ sensitivity ($k$): this allows us to add less noise.
Second, because the tails of the Gaussian noise distribution decay very fast, this makes the thresholding step more efficient in preventing zero counts from appearing in the output, keeping the spurious rate to acceptably low levels.
We quantify the privacy guarantees of the Gaussian mechanism using zero-concentrated DP~\cite{bun2016concentrated} (zCDP).

The third step is straightforward: all counts below a threshold $\tau$ are removed from the output.
This step is necessary because the first step produces many $<$page, country$>$ tuples for which the non-noisy user count is very low or even 0.
Such counts lead to unacceptable high relative error and spurious rate.
Conversations with data users showed that these made the output dataset hard to use, and that users were most interested in the most-viewed pages, rather than the long tail of pages with few views.
Suppressing counts below a fixed and configurable threshold $\tau$ addresses this problem, at the cost of a non-zero drop rate.

The mechanism is presented in Algorithm~\ref{algo:server-side}; in this algorithm,
$\mathcal{N}\left(0,\sigma^2\right)$ denotes a random sample from a normal distribution of mean 0 and variance $\sigma^2$. 
Step 1 uses only public data, Step 2 provides $\rho$-zCDP~\cite{bun2016concentrated}, and Step 3 is a post-processing step: the full algorithm satisfies $\rho$-zCDP.

\begin{algorithm}[htpb]
\caption{Server-side algorithm for the current pageviews}
\begin{algorithmic}[1]
\Require $t$: an ingestion threshold.
\Require $\tau$: a suppression threshold.
\Require $\rho$: a privacy parameter for zCDP.
\Require $P = \left<p_1,b_1\right>, \left<p_2,b_2\right>, \dots$: a private dataset of annotated pageviews, such each user is at most associated with $k$ unique pageviews $\pair{p_i,b_i}$ where $b_i=\texttt{true}$, and all of them have distinct $p_i$.
\Require $P_{daily} = \pair{p_1,n_1}, \pair{p_2,n_2}, \dots$: a public dataset listing the global number of pageviews for each page.
\Require $C$: a pre-defined list of countries.
\Statex \emph{Step 1: Collecting aggregation groups}
\State $G\leftarrow\{\}$
\For{$\pair{p,n}$ in $P_{daily}$}
    \If{$n \ge t$}
        \For{$c$ in $C$}
            \State $G \leftarrow G \cup \left<p,c\right>$
        \EndFor
    \EndIf
\EndFor
\Statex \emph{Step 2: Computing noisy counts}
\State $\sigma \leftarrow \sqrt{\frac{k}{2\rho}}$
\State $O \leftarrow \{\}$
\For{$g$ in $G$}
    \State $c \leftarrow \left|\left\{p\in P \mid p=g\right\}\right|$
    \State $\hat{c} \leftarrow c + \mathcal{N}\left(0,\sigma^2\right)$
    \State $O \leftarrow O \cup \left<g,\hat{c}\right>$
\EndFor
\Statex \emph{Step 3: Suppressing low counts}
\For{$\left<g,\hat{c}\right>$ in $G$}
    \If{$\hat{c} < \tau$}
        \State $O \leftarrow O \setminus \left<g,\hat{c}\right>$
    \EndIf
\EndFor
\State \Return $O$
\end{algorithmic}
\label{algo:server-side}
\end{algorithm}

We use $k=10$ as a per-user daily contribution bound, $t=150$ as an ingestion threshold, and $\tau=90$ as a suppression threshold.
These values were chosen after extensive experimentation, for input dataset completeness and to optimize the utility metrics described in Section~\ref{sec:accuracy}.

To select these algorithmic parameters, we computed metrics using the true data.
Such metrics are, in principle, sensitive, and the parameters themselves are not differentially private.
To mitigate the privacy risk from this tuning process, we kept fine-grained utility metrics confidential throughout the tuning process, minimizing data leakage. In addition to this consideration, we only publicly communicate approximate values of global utility metrics and the algorithmic parameters obtained from this tuning process.

Regardless, this remains a valid critique, and we would appreciate further research into the privacy loss entailed by confidentially tuning on sensitive metrics.

\subsection{Historical pageviews}\label{sec:historical}

To compute differentially private counts using the historical pageview dataset as input data, we follow a similar process, with one key difference: since the data is pre-aggregated, is is impossible to perform per-user contribution bounding.
Therefore, we do not use a client-side filtering step, and instead, use a different unit of privacy, as described in Section~\ref{sec:privacy-goal}.
We also have to sum the Count column of the pre-aggregated data, rather than simply counting the number of rows in each group.
Another difference is the use of Laplace noise instead of Gaussian noise, motivated by the fact that we only have a bound on the $L_1$ sensitivity of the aggregation, and not $L_2$ like with the current pageviews data.
The full process is otherwise similar to the previous one.
\begin{enumerate}
    \item First, we collect the list of $<$page, country$>$ tuples to aggregate over.
    \item Second, we sum the pageview counts in each group, and we add Laplace noise to each sum.
    \item Finally, we suppress low sums, and publish the data.
\end{enumerate}

The full algorithm is provided as Algorithm~\ref{algo:historical}; there, $\text{Lap}(0,\lambda)$ denotes a random sample from the Laplace distribution of mean 0 and scale $\lambda$.
Its privacy analysis is straightforward: Step 1 uses only public data, Step 2 provides $\eps$-DP guarantees~\cite{dwork2006calibrating}, and Step 3 is a post-processing step, so the full algorithm satisfies $\eps$-DP.

\begin{algorithm}[htpb]
\caption{Algorithm for the historical pageviews}
\begin{algorithmic}[1]
\Require $m$: the number of pageviews protected each day.
\Require $t$: an ingestion threshold.
\Require $\tau$: a suppression threshold.
\Require $\eps$: a privacy parameter for DP.
\Require $P_{hourly} = \pair{p_1,c_1}, \pair{p_2,c_2}, \dots$: a private dataset listing pre-aggregated hourly pageview counts.
\Require $P_{daily} = \pair{p_1,n_1}, \pair{p_2,n_2}, \dots$: a public dataset listing the global number of pageviews for each page.
\Require $C$: a pre-defined list of countries.
\Require $t$: A minimum pageview threshold for including pages in the output.
\Statex \emph{Step 1: Collecting aggregation groups} 
\State $G\leftarrow\{\}$
\For{$\pair{p,n}$ in $P_{daily}$}
    \If{$n \ge t$}
        \For{$c$ in $C$}
            \State $G \leftarrow G \cup \left<p,c\right>$
        \EndFor
    \EndIf
\EndFor
\Statex \emph{Step 2: Computing noisy sums}
\State $\lambda \leftarrow \frac{m}{\eps}$
\State $O \leftarrow \{\}$
\For{$g$ in $G$}
    \State $s \leftarrow \sum_{\pair{p,c} \in P_{hourly} \text{where} p=g} c$
    \State $\hat{s} \leftarrow s + \text{Lap}\left(0,\lambda\right)$
    \State $O \leftarrow O \cup \left<g,\hat{s}\right>$
\EndFor
\Statex \emph{Step 3: Suppressing low counts}
\For{$\left<g,\hat{s}\right>$ in $G$}
    \If{$\hat{s} < \tau$}
        \State $O \leftarrow O \setminus \left<g,\hat{s}\right>$
    \EndIf
\EndFor
\State \Return $O$
\end{algorithmic}
\label{algo:historical}
\end{algorithm}

As mentioned in Section~\ref{sec:privacy-goal}, we use $m=300$ for the 2015--2017 data, and $m=30$ for the 2017--2023 data.
For the 2015--2017 data, we use $t=150$ as ingestion threshold and $\tau=3500$ as suppression threshold. For the 2017--2023 data, we use $t=150$ as ingestion threshold and $\tau=450$ as suppression threshold.
These values were chosen to optimize the global utility metrics described in Section~\ref{sec:accuracy}.

\subsection{Implementation}\label{sec:implementation}

The algorithms were implemented and deployed using Tumult Analytics~\cite{tumultanalyticssoftware,tumultanalyticswhitepaper}, a framework chosen for its robustness, production-readiness, compatibility with Wikimedia's compute infrastructure, and support for advanced features like zCDP-based privacy accounting~\cite{wmfdpcomparison}.
This incurs very slight differences in the mechanisms used: on integer-valued data, Tumult Analytics uses a two-sided geometric distribution instead of Laplace noise, and a discrete version of the Gaussian mechanism~\cite{canonne2020discrete}.
The data release based on the current input data required implementing a new notion of neighboring relation in the framework: rather than protecting a fixed number of rows, or an arbitrary number of rows associated with a single user identifier, it protects a fixed number of rows \emph{associated with different aggregation groups}.
This was made easier by the extensibility of the underlying framework, Tumult Core.

\section{Outcomes}\label{sec:outcomes}

The deployment of this differentially private data publication project is now allowing the Wikimedia Foundation to release a much larger and much richer dataset about user visits to Wikimedia projects.
The magnitude of this increase in published pageview data is summarized in Table~\ref{tab:outcomes}.

\begin{table}[htb]
    \centering
    \begin{tabular}{c|c|c|c}
         & Before this project & After this project & Percentage change \\
         \hline
         \begin{tabular}{@{}c@{}} Median number of data points \\ released per day \end{tabular}
         & 9,000 & 360,000 & +4,000\% \\
         \hline
         \begin{tabular}{@{}c@{}} Median number of pageviews \\ released per day \end{tabular}
         & 50 million & 120 million & +240\% \\
         \hline
         \begin{tabular}{@{}c@{}} Total number of data points \\ released since 2021 \end{tabular}
         & 8 million & 120 million & +1,500\% \\
         \hline
         \begin{tabular}{@{}c@{}} Total number of pageviews \\ released since 2021 \end{tabular}
         & 47 billion & 116 billion & +250\% \\
    \end{tabular}
    \caption{\label{tab:outcomes}A comparison of the amount of data published before and after this project, as of June 29, 2023.}
\end{table}

More than 2,000 days of historical data from 2015 to 2021 were not previously published.
The use of differential privacy in this project allowed the Wikimedia Foundation to release more than 135 million statistics about this data, encompassing 325 billion pageviews.

The output data had acceptable quality according to our success metrics.

\begin{itemize}
    \item For the data based on the current pageviews dataset, more than 95\% of the counts has a relative error below 50\%, the drop rate above 150 is below 0.1\%, the global spurious rate is below 0.01\%, and below 3\% for all but 3 countries.
    \item For the 2017--2023 data, the median top-1000 drop rate is below 8\%, the drop rate above 450 is below 3\%, and the global spurious rate is below 0.1\%.
    \item For the 2015--2017 data, the top-1000 drop rate is below 40\%, the drop rate above 3500 is below 3\%, and the global spurious rate is below 20\%.
\end{itemize}

These metrics show that the privacy-accuracy trade-offs are much better for recent data than for historical data: this is explained by the much tighter sensitivity bound from client-side filtering, allowing to take full advantage of the Gaussian mechanism and its fast-decaying tails.

\section{Conclusion}

In this paper, we described the process and mechanisms that allowed the Wikimedia Foundation to publish large-scale datasets about user behavior on Wikipedia and other Wikimedia projects.
Multiple key factors made this launch possible.

\begin{itemize}
\item Tumult Labs' systematic workflow for differential privacy publications, described in Section~\ref{sec:workflow}, provided the structure necessary to move the project forward from its inception to its deployment.
\item Combining client-side filtering with server-side aggregation, as described in Section~\ref{sec:current}, was a key innovation that allowed us to obtain user-level differential privacy guarantees for the current pageview data without tracking user identifiers.
\item Tumult Core, the privacy framework underlying Tumult Analytics, is designed for extensibility. This made it possible for us to add a novel neighboring definition to this framework to capture the properties of client-side filtering, while still being able to use tight privacy accounting techniques.
\item Finally, the scalability offered by Tumult Analytics was essential in handling the massive datasets that were used as input in this project.
\end{itemize}

The data is now published online~\cite{wmfreadmecurrent,wmfreadmehistorical,wmfreadmepre2017}, along with the source code of the client-side filtering infrastructure~\cite{wmfcookiecode} and the server-side algorithms~\cite{wmfdpcode,wmfdpcodehistorical}.
We look forward to seeing what use cases this data will enable!
\section{Acknowledgements}

We are grateful to Luke Hartman, Tomoko Kitazawa, Nuria Ruiz, and Xabriel J. Collazo Mojica for their help with this project, and to Leila Zia and the anonymous reviewers for their helpful comments and suggestions on this paper.

\bibliographystyle{plain}
\bibliography{biblio.bib}

\end{document}